\documentclass[prl,aps,nofootinbib,notitlepage,twocolumn]{revtex4-2}

\usepackage{amsthm}
\usepackage{amsmath,bm}
\usepackage{amssymb}
\usepackage{amsfonts}
\usepackage{graphicx}

\usepackage{txfonts}

\usepackage[colorlinks=true,linkcolor=blue,citecolor=blue,urlcolor=blue]{hyperref}

\begin{document}

\title{Scaling laws for the sensitivity enhancement of non-Gaussian spin states}
\author{Youcef Baamara}
\email{youcef.baamara@lkb.ens.fr}
\author{Alice Sinatra}
\email{alice.sinatra@lkb.ens.fr}
\author{Manuel Gessner}
\email{manuel.gessner@ens.fr}
\affiliation{Laboratoire Kastler Brossel, ENS-Universit\'{e} PSL, CNRS, Sorbonne Universit\'{e}, Coll\`{e}ge de France, 24 Rue Lhomond, 75005, Paris, France}
\date{\today}

\begin{abstract}
We identify the large-$N$ scaling of the metrological quantum gain offered by over-squeezed spin states that are accessible by one-axis twisting, as a function of the preparation time. We further determine how the scaling is modified by relevant decoherence processes and predict a discontinuous change of the quantum gain at a critical preparation time that depends on the noise. Our analytical results provide recipes for optimal and feasible implementations of quantum enhancements with non-Gaussian spin states in existing experiments, well beyond the reach of spin squeezing.
\end{abstract}

\maketitle

\textit{Introduction.---}Quantum projection noise limits the precision of quantum measurements, even after all external noise sources have been eliminated. The so-called standard quantum limit (SQL) defines the maximal precision of classical measurement strategies based on coherent states, and can be overcome by using nonclassical quantum states and suitable measurement observables~\cite{CavesPRD1981,WinelandPRA1992,BollingerPRA1996,GiovannettiNATPHOT2011,PezzeRMP2018}. In Ramsey spectroscopy measurements, the central element of atomic clocks and interferometers, the uncertainty for the estimation of an unknown phase parameter $\theta$ at the SQL is given by $(\Delta\theta)_{\mathrm{SQL}}^2=1/N$, where $N$ is the total number of atoms. Quantum strategies, in principle, allow for quantum enhancements as large as $(\Delta\theta)_{\mathrm{SQL}}^2/(\Delta\theta)^2=N$, but reaching this so-called Heisenberg limit (HL) involves the generation of highly entangled many-body states that require long preparation times and are prone to decay quickly under decoherence~\cite{HuelgaPRL1997,MonzPRL2011,DemkowiczNATCOMMUN2012,Kittens}. Instead, an experimentally more robust approach consists in the squeezing of collective spin observables on shorter timescales, in close analogy to the quantum optical squeezing~\cite{CavesPRD1981,KimblePRL1987,GrangierPRL1987} that improves the sensitivity of gravitational wave detectors~\cite{TsePRL2019}. Atomic spin squeezing, nowadays routinely produced in the laboratory, has been shown to enhance proof-of-principle interferometric measurements~\cite{GrossNATURE2010,MitchellPRL2012,HostenNATURE2016,CoxPRL2016,PezzeRMP2018}, and to reveal many-particle entanglement~\cite{SorensenNATURE2001,SMPRL01,EsteveNATURE2008,RiedelNATURE2010,LerouxPRL2010,TothJPA2014,RenPRL2021}.

Starting from an uncorrelated ensemble, quantum entanglement can be generated by means of a nonlinear evolution, such as one-axis twisting~\cite{KitagawaPRA1993,PezzeRMP2018}.
The sensitivity enhancement that is offered by spin squeezing in this case is fundamentally limited to a gain of $(\Delta\theta)_{\mathrm{SQL}}^2/(\Delta\theta)^2\sim N^{2/3}$~\cite{KitagawaPRA1993}. This limitation arises because of the restriction to the measurement of linear spin observables, which no longer capture the fine features of non-Gaussian quantum states~\cite{GessnerPRL2019}. A recording of the full measurement statistics may overcome this problem, as can be shown by an analysis of the Fisher information~\cite{PezzePRL2009,PezzeRMP2018}, but this is often unpractical due to the demanding requirements in measurement resolution. An alternative is given by measurement-after-interaction strategies (MAI), such as squeezing echos, which combine nonlinear evolutions with the measurement of linear observables~\cite{YurkePRA1986,SchleierSmithPRL2016,FrowisPRL2016,MacriPRA2016,HostenSCIENCE2016,NolanPRL2017,HammererQuantum2020}. These strategies can in principle achieve Heisenberg scaling, i.e., a quantum enhancement proportional to the atom number $N$, even in the presence of considerable detection noise~\cite{SchleierSmithPRL2016,FrowisPRL2016,NolanPRL2017}. However, this requires a stable and coherent interacting evolution on sufficiently long timescales, which remains a challenge in systems with large $N$. 

One-axis twisting generates non-Gaussian spin states with sensitivities beyond the reach of spin-squeezed states already on much shorter timescales. Such over-squeezed spin states are created after an evolution that is longer than that of spin-squeezed states ($\chi t\sim N^{-2/3}$, where $\chi$ is the frequency associated with the nonlinearity in the one-axis-twisting Hamiltonian), but remains far below the times ($\chi t\sim N^{-1/2}$) that are required to reach Heisenberg scaling. States in this intermediate regime are already available with considerable particle numbers in state-of-the-art experiments~\cite{StrobelSCIENCE2014,BohnetSCIENCE2016} and are of principal interest for experimental demonstrations of metrological quantum enhancements beyond squeezing~\cite{EvrardPRL2019,XuArXiv2021}. The non-Gaussian fluctuations of these states, however, prevent a practical characterization of their properties with standard methods, such as the spin-squeezing coefficient~\cite{WinelandPRA1992}, and consequently their metrological potential in the relevant limit of large $N$ has so far remained unknown.

In this article, we analytically determine the scaling of quantum enhancements for states generated by one-axis-twisting dynamics on timescales $\chi t\sim N^{-\alpha}$ with $1\geq \alpha\geq 1/2$. We show that suitable nonlinear observables for non-Gaussian spin states can be efficiently accessed by MAI techniques, leading to a maximal quantum gain that scales as $(\Delta\theta)_{\mathrm{SQL}}^2/(\Delta\theta)^2\sim N^{2-2\alpha}$, where $\alpha$ determines the preparation time. We demonstrate that this scaling law fully describes the metrological potential of non-Gaussian over-squeezed states and continuously connects various optimal strategies spanning from the SQL ($\alpha=1$), linear ($\alpha=2/3$) and second-order squeezing ($\alpha=4/7$), up to the Heisenberg scaling of echo protocols on longer timescales ($\alpha=1/2$). Finally, we study how this scaling is modified by dominant dephasing processes in atomic experiments. In the thermodynamic limit $N\gg 1$, we analytically predict a discontinuous change in the scaling of the quantum gain at a critical value of $\alpha$, which naturally determines the optimal nonlinear evolution time as a function of the dephasing rate. In realistic atomic systems, this abrupt transition is washed out by finite-size effects but the onset of a discontinuity can be observed as the atom number is increased. 

\textit{From linear to second-order spin squeezing.---}The goal of an atomic clock is to determine with high precision the energy difference $\hbar\omega$ between two internal atomic levels. The information about $\omega$ is contained in the rotation angle $\theta=\omega T$ of the collective spin state of the atoms after an evolution time $T$. We consider an ensemble of $N$ atoms that is described by collective spin observables $\hat{\vec{S}}=(\hat{S}_x,\hat{S}_y,\hat{S}_z)^T$, with $\hat{S}_{k}=\sum_{i=1}^N\hat{\sigma}^{(i)}_k/2$ and $\sigma^{(i)}_k$ is a Pauli matrix for the $i$th atom. The axis $\vec{n}$ of the rotation $e^{-i\hat{S}_{\vec{n}}\theta}$ with $\hat{S}_{\vec{n}}=\vec{n}\cdot\hat{\vec{S}}$ can be controlled by rotations of the spin state with external fields. To estimate the value of $\theta$, the average value of an observable $\hat{X}$ is measured, which, after many repeated measurements, leads to an estimation error of $(\Delta \theta)^2=(\Delta\hat{X})^2/|\langle[\hat{X},\hat{S}_{\vec{n}}]\rangle|^2$~\cite{WinelandPRA1992,BollingerPRA1996}.

Spin-coherent states yield the lowest achievable estimation error among all separable spin states, which determines the SQL as $(\Delta \theta)_{\mathrm{SQL}}^2=1/N$. Entangled spin states with quantum fluctuations below the SQL can achieve lower uncertainties $(\Delta\theta)^2=\xi^{2}/N$, where $\xi<1$ indicates a quantum enhancement that is quantified by the parameter $\xi^{-2}=(\Delta \theta)_{\mathrm{SQL}}^2/(\Delta\theta)^2=|\langle[\hat{X},\hat{S}_{\vec{n}}]\rangle|^2/(N(\Delta\hat{X})^2)$~\cite{WinelandPRA1992,GessnerPRL2019}. To maximize the sensitivity, the rotation axis $\vec{n}$ and the measurement observable $\hat{X}$ are optimized, under the consideration of possible constraints. If there are no constraints on the measurement observable $\hat{X}$, the full metrological potential of the state can be harnessed, leading to the maximal sensitivity gain $\max_{\hat{X}}\xi^{-2}=F_Q/N$, where $F_Q$ is the quantum Fisher information~\cite{BraunsteinPRL1994}.

For the generation of metrologically useful entangled states, we consider the one-axis-twisting evolution $|\psi(t)\rangle=e^{-i\hat{H} t/\hbar}|\psi_{0}\rangle$ of an initial spin-coherent state $|\psi_0\rangle$ with $\hat{S}_x|\psi_0\rangle=\frac{N}{2}|\psi_0\rangle$, generated by the nonlinear Hamiltonian $H=\hbar \chi \hat{S}_z^2$. On short times, this leads to the generation of spin-squeezed states whose features are efficiently captured by linear (L) spin observables $\hat{X}=\hat{S}_{\vec{m}}$ and in the limit $N\gg 1$ achieve the quantum enhancement $\xi_{\mathrm{L}}^{-2} \simeq 2\times 3^{-2/3}N^{2/3}$~\cite{KitagawaPRA1993,SinatraFro2012,footnotesimeq}. An evolution beyond the optimal squeezing time $\chi t\simeq 3^{1/6} N^{-2/3}$ gives rise to non-Gaussian spin states whose metrological sensitivity is higher, but can only be extracted through the measurement of nonlinear spin observables, which can be optimized~\cite{GessnerPRL2019}.

\begin{figure}[tb]
     \centering
     \includegraphics[width=0.48\textwidth]{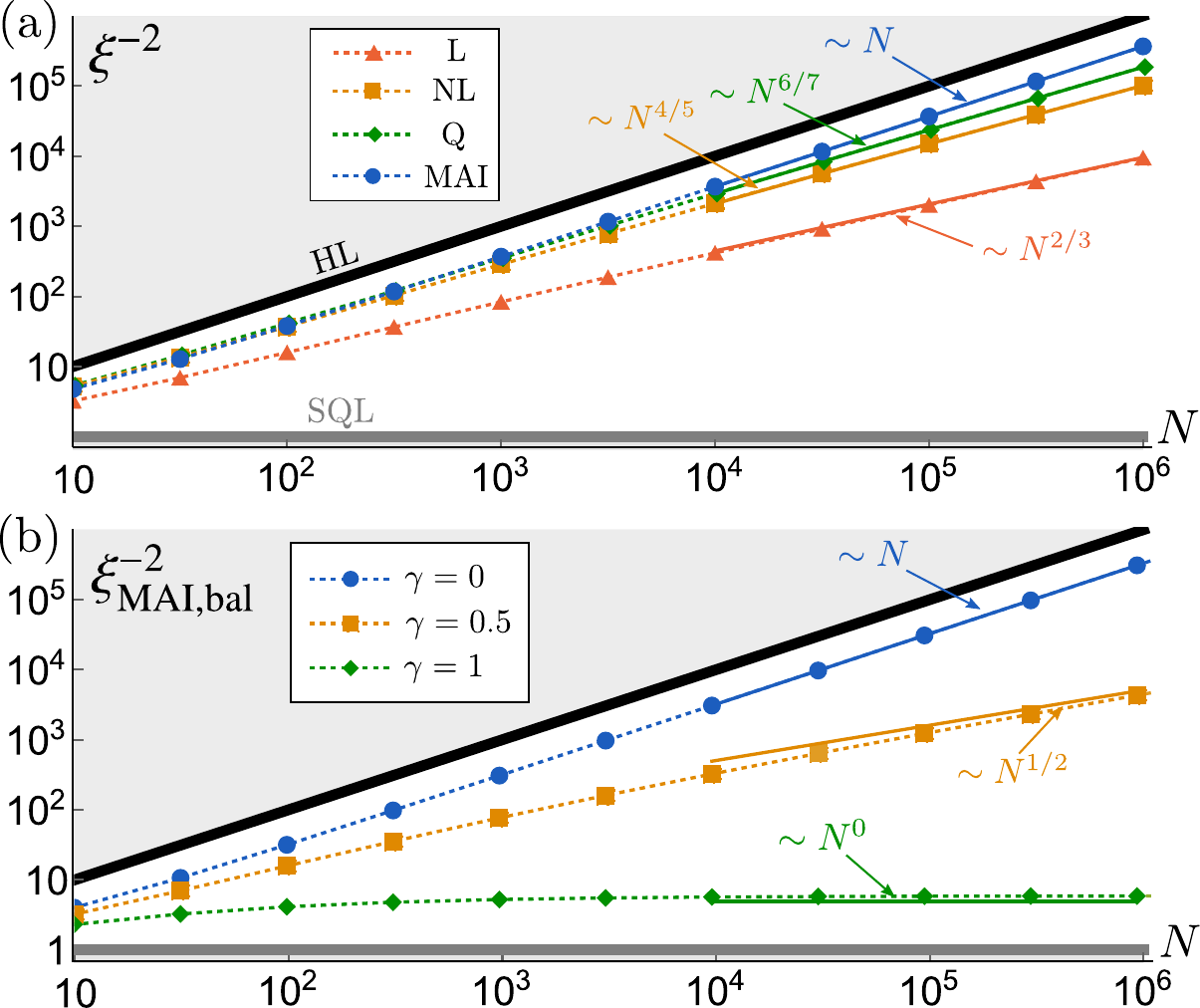}
     \caption{\textbf{Scaling laws for the quantum gain.} (a) The quantum metrological gain $\xi^{-2}$ optimized over the preparation time $\chi t$ for linear (L), nonlinear (NL), quadratic (Q), and MAI measurement strategies as a function of $N$ compared to the standard quantum limit (SQL) and the Heisenberg limit (HL). Solid lines represent the analytical scaling laws in the limit of large $N$ including finite size corrections. The leading term is indicated in the plot. (b) Modification of the scaling laws for the MAI method in the presence of ballistic noise with fluctuations that scale with $\gamma$ for $\epsilon=0.05$ (see text).}
     \label{scaling}
 \end{figure}

Interestingly, already the measurement of a single second-order observable $\{\hat{S}_x,\hat{S}_z\}$, where $\{\hat{A},\hat{B}\}=\hat{A}\hat{B}+\hat{B}\hat{A}$ is the anticommutator, in addition to $\hat{S}_{\vec{m}}$, enables a significant gain over linear spin squeezing. An analytical calculation of moments up to fourth order of the collective spin components of the state $|\psi(t)\rangle$ for arbitrary $N$~\cite{LongPaper} allows us to use the technique of Ref.~\cite{GessnerPRL2019} to identify the optimal nonlinear (NL) measurement observable of the form $\hat{X}_{\mathrm{NL}}=m_y \hat{S}_y+m_z \hat{S}_z +\frac{m_{xz}}{2} \{\hat{S}_x,\hat{S}_z\}$, as well as the optimal rotation axis $\vec{n}$, such that the sensitivity for the states $|\psi(t)\rangle$ is maximized. In the limit $N\gg 1$, we find that optimal over-squeezed states are generated after $\chi t\simeq (5/2)^{1/10} 3^{3/10}N^{-3/5}$ and yield a maximum quantum gain of
\begin{equation}\label{xi3}
\xi_{\mathrm{NL}}^{-2}\simeq 2 \left(\frac{2}{5}\right)^{4/5} 3^{3/5} N^{4/5}.
\end{equation}
We further obtain the leading corrections for finite-sized systems as $\xi^{-2}_{\mathrm{NL},N}=\xi_{\mathrm{NL}}^{-2}[1-(5/2)^{1/5} 3^{3/5} N^{-1/5}+\frac{893}{2^{2/5}3^{4/5}5^{3/5}28}N^{-2/5}+\mathcal{O}(N^{-3/5})]$. Extending this optimization to the measurement of arbitrary linear combinations of spin observables up to second order yields the maximal quantum gain for quadratic (Q) spin squeezing $\xi_{\mathrm{Q}}^{-2}\simeq (6/7)^{6/7}5^{3/7} N^{6/7}$ after a preparation time $\chi t\simeq (7/6)^{1/14} 5^{3/14}N^{-4/7}$~\cite{footnoteQ}. The analytical scaling laws for linear, nonlinear and second-order spin squeezing are compared in Fig.~\ref{scaling}~(a).

Nonlinear spin observables can be extracted from the higher moments of the measurement statistics of linear spin observables~\cite{LuckeSCIENCE2011,StrobelSCIENCE2014,BohnetSCIENCE2016,EvrardPRL2019}. However, this increases the measurement time and requires low detection noise, which is challenging to achieve in systems with large atom numbers. This requirement can be avoided by MAI techniques that employ an additional nonlinear evolution before the measurement of a linear spin observable~\cite{SchleierSmithPRL2016,FrowisPRL2016,NolanPRL2017}. To identify the possibilities of this technique in the present context, we consider a second one-axis-twisting evolution $\hat{U}_{\tau}=e^{-i\chi\tau\hat{S}_z^2}$, which is implemented after the phase imprinting and is followed by a measurement of $\hat{S}_{\vec{m}}$. This effectively gives access to the nonlinear observables $\hat{X}_{\mathrm{MAI}}=\hat{U}_{\tau}^{\dagger} \hat{S}_{\vec{m}}\hat{U}_{\tau}$ or
\begin{align}\label{eq:XMAI}
\hat{X}_{\mathrm{MAI}}=e^{-i\chi \tau}\left(\cos(2 \chi \tau\hat{S}_z)\hat{S}_{\vec{m}_1}+\sin(2 \chi \tau\hat{S}_z)\hat{S}_{\vec{m}_{2}}\right)+m_z\hat{S}_z,
\end{align}
where $\vec{m}_1=(m_x,m_y,0)^T$, $\vec{m}_{2}=(m_y,-m_x,0)^T$. For small $\tau$, we can expand~(\ref{eq:XMAI}) up to linear order in $\tau$, and with $m_x=0$, we obtain the nonlinear observable $\hat{X}_{\mathrm{MAI}}=\hat{S}_{\vec{m}}+\chi \tau m_y \{\hat{S}_x,\hat{S}_z\}+\mathcal{O}(\chi \tau)^2$. Hence, a nonlinear evolution up to $\chi\tau=m_{xz}/(2 m_y)$ followed by a measurement of the collective spin observable $\hat{S}_{\vec{m}}$ with $\vec{m}=(m_x,m_y,m_z)^T$ is equivalent to first order in $\tau$ to the measurement of the optimal observable $\hat{X}_{\mathrm{NL}}$ for intermediate over-squeezed states~\cite{footnoteSz}.

\textit{Scaling laws for MAI methods.---}We now explore the general potential of the continuous set of observables~(\ref{eq:XMAI}) that are made available by MAI techniques for the states $|\psi(t)\rangle$. To this end, we analytically identify the optimal choice of $\vec{m}\in\mathbb{R}^3$ in $\hat{X}_{\mathrm{MAI}}$ that maximizes the sensitivity gain $\xi^{-2}$ for arbitrary $t$ and $\tau$. In the next step, numerical optimizations over $\tau$ reveal that in the limit of large $N$, the choice $\tau=-t$ maximizes the sensitivity gain in the relevant time frame $\chi t\lesssim 1/\sqrt{N}$~\cite{Hammererfootnote}. This corresponds to the echo protocol that was first suggested in Ref.~\cite{SchleierSmithPRL2016} to achieve Heisenberg scaling with long state preparation times. 
In the following we identify the metrological potential of the echo protocol with $\tau=-t$ for arbitrary states that are prepared on the timescales $\chi t = \sigma N^{-\alpha}$ with $1\geq \alpha\geq 1/2$. We obtain the scaling law for the optimal quantum gain in the limit $N\gg 1$, as a function of the preparation timescale, as
\begin{align}\label{MAIopt}
\xi^{-2}_{\mathrm{MAI}}\simeq
\begin{cases}
\sigma^2 N^{2-2\alpha}, &\quad  1\geq \alpha > 1/2\\
\sigma^2e^{-\sigma^2} N, &\quad \alpha = 1/2
\end{cases}.
\end{align}

We first note that the scaling law~(\ref{MAIopt}) reproduces and continuously interpolates between all the cases that were discussed above: At relatively short times, $\alpha=2/3$, we recover the quantum gain $\xi^{-2}_{\mathrm{L}}\sim N^{2/3}$ of (linear) spin-squeezed states. For $\alpha=3/5$, we recover the $\xi^{-2}_{\mathrm{NL}}\sim N^{4/5}$ scaling of intermediate, over-squeezed states, and for $\alpha=4/7$ the full potential of second-order spin squeezing, $\xi^{-2}_{\mathrm{Q}}\sim N^{6/7}$, is exploited. Maximal sensitivity is obtained when the nonlinear evolution can be implemented on the longest possible timescales, i.e., at $\alpha=1/2$. In this case, after an additional optimization over $\sigma$, we recover the Heisenberg scaling $\xi^{-2}_{\mathrm{MAI}}=N/e$ for states at $\chi t=1/\sqrt{N}$~\cite{SchleierSmithPRL2016}. As we will discuss in further detail below, in practical situations the minimal achievable $\alpha$ is limited by decoherence.

The result~(\ref{MAIopt}) shows that the MAI technique effectively uncovers the metrological features of squeezed and non-Gaussian spin states over a wide range of timescales. To fully assess the quality of the sensitivity~(\ref{MAIopt}) that is extracted by MAI, we compare to the sensitivity that can be achieved by the unconstrained optimization over all possible quantum observables $\hat{X}$ to capture the features of the states $|\psi(t)\rangle$, leading to the quantum Fisher information. This optimization can be carried out analytically and yields in the limit $N\gg 1$:
\begin{align}\label{Fopt}
\max_{\hat{X}}\xi^{-2}=F_Q/N\simeq
\begin{cases}
\sigma^2 N^{2-2\alpha}, & 1\geq \alpha > 1/2\\
\frac{1}{2}(1-e^{-2\sigma^2}) N, & \alpha = 1/2
\end{cases}.
\end{align}
Comparison with~(\ref{MAIopt}) reveals the remarkable optimality of the MAI protocol over the entire range of timescales $1\geq \alpha > 1/2$. Even though this protocol requires only measurements of the average value of the observable~(\ref{eq:XMAI}), no other measurement scheme could achieve higher sensitivity based on the same quantum state $|\psi(t)\rangle$. At $\alpha=1/2$, MAI yields Heisenberg scaling with a modified pre-factor.

\textit{Effect of decoherence.---}Our results so far predict an increasing quantum gain with longer evolution times. In order to determine the limitations on the coherent nonlinear evolution times, we now consider the impact of dominant decoherence processes on these scaling laws. Different realizations of the one-axis-twisting Hamiltonian are dominated by different decoherence processes. Implementations based on cold collisions in Bose-Einstein condensates~\cite{GrossNATURE2010,SorensenNATURE2001,EPJD,RiedelNATURE2010} are fundamentally limited by particle losses and finite temperature~\cite{LiYunPRL2008,SinatraPRL2011}, whose effect on spin squeezing can both be described with a dephasing model and lead to ballistic behavior of the spin fluctuations $(\Delta \hat{S}_y)^2$~\cite{SinatraFro2012}. Similarly, in trapped-ion implementations~\cite{MolmerPRL1999,MonzPRL2011,BohnetSCIENCE2016}, ballistic collective dephasing is caused by fluctuating magnetic fields~\cite{MonzPRL2011,LanyonPRL2013,CarnioPRL2015}. Another path is offered by cavity squeezing of cold thermal atomic ensembles~\cite{LerouxPRL2010}. In this case, cavity loss defines the dominant noise process, and is described by a collective dephasing of the spin that is of diffusive nature~\cite{LerouxPRA2012, PawlowskiEPL2016}.

\textit{Ballistic dephasing.---}We first focus on the case of a time-independent, random dephasing process. This process is governed by the evolution of the Hamiltonian $\hat{H}=\hbar \chi (\hat{S}_z^2 + D \hat{S}_z)$, where a phase rotation of constant strength $D$ has been added to the one-axis-twisting Hamiltonian. We assume that $D$, which fluctuates from one experimental realization to the next, is a Gaussian random variable with $\langle D\rangle=0$ and variance $\langle D^2 \rangle = \epsilon  N^\gamma$, with $0 \leq \gamma \leq 1$. Starting from the spin-coherent state $|\psi_0\rangle$ and considering the MAI scheme with two nonlinear evolutions of duration $t$, the spin moments under this noise process can be determined analytically, showing that in the limit $N\gg 1$, the noise in the relevant spin observable increases as $\frac{(\Delta \hat{S}_y)^2}{N/4}\simeq 1+4\epsilon N^{1+\gamma}(\chi t)^2+\mathcal{O}(\chi t)^4$. The characteristic quadratic increase of the variance over time reveals the ballistic nature of this dephasing process. By including the effect of ballistic dephasing in our optimization of the quantum gain, we observe that the scaling law~(\ref{MAIopt}) is modified to
\begin{align}\label{eq:MAIb}
\xi^{-2}_{\mathrm{MAI,bal}}\simeq
\begin{cases}
    \frac{\sigma^2 N^{2-2\alpha}}{1+4\epsilon \sigma^2 N^{1+\gamma-2\alpha}}, &\: 1\geq \alpha > 1/2\\
\frac{\sigma^2e^{-\sigma^2}N}{1+4\epsilon\sigma^2 N^{\gamma}}, &\: \alpha=1/2
\end{cases}.
\end{align}
The scaling with $N$ now depends on the precise interplay of the evolution time and the noise via $\alpha$ and $\gamma$, see Fig.~\ref{scaling}~(b). We observe an abrupt change of the scaling law at the critical value of
\begin{align}\label{eq:alpha_c}
\alpha_c=\frac{1+\gamma}{2}.
\end{align}
For times shorter than this critical value, $\alpha>\alpha_c$, we obtain
\begin{align}\label{eq:MAIbshort}
\xi^{-2}_{\mathrm{MAI,bal}} \simeq \sigma^2 N^{2-2\alpha},
\end{align}
i.e., the scaling~(\ref{MAIopt}) of the ideal, noiseless evolution is preserved. For longer times, $\alpha < \alpha_c$, however, the quantum gain
\begin{align}\label{eq:MAIblong}
\xi^{-2}_{\mathrm{MAI,bal}}\simeq
 \frac{1}{4\epsilon } N^{1-\gamma}
\end{align}
is limited by the dephasing process. Note that the long-time scaling always represents the maximum quantum gain for any given $\gamma$, i.e., $1-\gamma\geq 2-2\alpha$. This scaling is achieved exactly at the critical point $\alpha_c$, as well as by all longer times since~(\ref{eq:MAIblong}) is independent of $\alpha$. The critical value~(\ref{eq:alpha_c}) thus naturally identifies the optimal preparation time $\alpha=\alpha_c$ as a function of the dephasing strength $\gamma$ in the thermodynamic limit $N\to \infty$. Furthermore, at $\alpha=\alpha_c$, the results~(\ref{eq:MAIbshort}) and~(\ref{eq:MAIblong}) predict a discontinuous change of the scaling law as a function of the preparation time. 

\begin{figure}[tb]
    \centering
    \includegraphics[width=.48\textwidth]{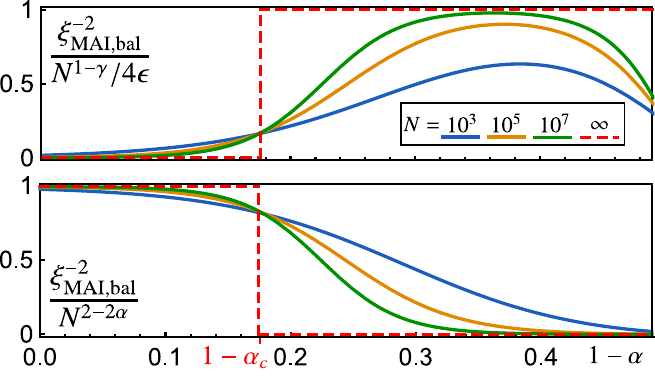}
    \caption{\textbf{Discontinuous scaling law in the presence of ballistic dephasing.} Quantum metrological gain of the MAI method $\xi^{-2}_{\mathrm{MAI,bal}}$ with $\tau=-t$, $\epsilon=0.05$, $\gamma=0.65$ as a function of $\alpha$: Larger values of $1-\alpha$ correspond to longer preparation times as we set $\chi t=\sigma N^{-\alpha}$. The normalized quantum gain $\xi^{-2}_{\mathrm{MAI,bal}}/(N^{1-\gamma}/4\epsilon)$ (top) and $\xi^{-2}_{\mathrm{MAI,bal}}/N^{2-2\alpha}$ (bottom) illustrates how the scaling in the thermodynamic limit, Eqs.~(\ref{eq:MAIbshort}) and~(\ref{eq:MAIblong}) for short and longer times, respectively, is approached as $N$ increases. At $\alpha_c$~(\ref{eq:alpha_c}) the transition becomes discontinuous in the thermodynamic limit.}
    \label{Deph}
\end{figure}
This critical behavior is blurred by finite-size effects in realistic atomic samples, but an onset of the discontinuity is discernible with increasing $N$, see Fig.~\ref{Deph}. Higher-order corrections to Eq.~(\ref{eq:MAIb}) show that  $\xi^{-2}_{\mathrm{MAI,bal}}$ exhibits a maximum in the region $\alpha<\alpha_c$ at $\chi t\simeq(4\epsilon)^{-1/4}N^{-1/2-\gamma/4}$ where the optimal sensitivity~(\ref{eq:MAIblong}) is attained. This maximum becomes increasingly flat as $N$ grows until the maximum sensitivity~(\ref{eq:MAIblong}) is attained by any value of $\alpha\leq\alpha_c$ in the thermodynamic limit.

In the limit $\gamma=0$, we find that the metrological quantum gain is entirely insensitive to decoherence, independently of $\alpha$. In general, we can interpret the critical value~(\ref{eq:alpha_c}) as a definition of the maximal tolerable level of ballistic dephasing noise $\gamma=2\alpha-1$ as a function of the desired quantum gain, which scales with $\alpha$. Naturally, these constraints become more demanding for higher quantum enhancements. For instance, to achieve the optimal sensitivity enhancement of linear spin squeezing $\xi^{-2}_{\mathrm{L}}\sim N^{2/3}$, noise with $\gamma\leq 1/3$ may be tolerated in the experiment, whereas second-order spin squeezing with $\xi^{-2}_{\mathrm{Q}}\sim N^{6/7}$ requires $\gamma\leq 1/7$. In the extreme case $\gamma=1$, our result confirms that strong dephasing prevents the generation of quantum enhancements that increase with $N$~\cite{SinatraFro2012}.

It was shown in Ref.~\cite{SinatraFro2012} that the effect of particle losses on optimal linear spin squeezing is well described by ballistic dephasing with $\gamma=1$ and $\epsilon=N_{\mathrm{loss}}/(3N)$, where $N_{\mathrm{loss}}$ is the average number of lost atoms. This correspondence can indeed be generalized to nonlinear spin squeezing, $\xi^{-2}_{\mathrm{NL}}$, as we have confirmed analytically using Monte-Carlo wave function methods, extending the results of Ref.~\cite{LiYunPRL2008} to the case of a nonlinear measurement. This implies that for a given atom number, the nonlinear measurement leads to an improvement over linear spin squeezing as long as $N/N_{\mathrm{loss}}\gtrsim N^{2/3}$.

\textit{Diffusive dephasing.---}Finally, we consider a dephasing process described by the Lindblad master equation $(\partial/\partial t) \hat{\rho}=-i[\hat{H},\hat{\rho}]+\gamma_C(\hat{S}_z\hat{\rho} \hat{S}_z+\frac{1}{2}\{\hat{S}_z^2,\rho\})$, where the relative strength of the dephasing is quantified by the parameter $\epsilon=\gamma_C/\chi$. In contrast to the previous noise process, here the variance $\frac{(\Delta \hat{S}_y)^2}{N/4}\simeq 1+ 2\epsilon N \chi t+\mathcal{O}(\chi t)^2$ increases linearly in time, revealing the diffusive nature of this dephasing processes. An analogous procedure now yields for any $\epsilon>0$:
    \begin{align}
    \xi^{-2}_{\mathrm{MAI,dif}}\simeq
    \begin{cases}
    \frac{\sigma}{2\epsilon}N^{1-\alpha}, &\quad 1\geq \alpha > 1/2\\
    \frac{\sigma e^{-\sigma^2}}{2\epsilon}N^{1/2}, &\quad \alpha = 1/2
    \end{cases}.
\end{align}
Diffusive dephasing thus leads to a reduction of the scaling exponent by a factor of $2$, independently of the preparation time $\alpha$. An optimization over $\alpha$ and $\sigma$ shows that, at fixed $\epsilon$, the largest achievable quantum gain is given by $\xi^{-2}\sim N^{1/2}$, which analytically confirms a result that was numerically obtained in Ref.~\cite{HammererQuantum2020}. When choosing $\alpha=3/5$, we obtain a maximal gain of $\xi^{-2} \sim N^{2/5}$. This scaling was found to be optimal for linear spin squeezing in the presence of diffusive dephasing due to cavity losses
\cite{PawlowskiEPL2016,MonikaPRA2010,LerouxPRA2012}.

\textit{Unifying expression.---}Our results can be summarized in a single unifying expression for the sensitivity gain in the limit of large $N$ that is valid over the full range of timescales $\chi t<N^{-1/2}$:
\begin{align}\label{eq:unifying}
    \xi^{-2}\simeq \frac{N^2(\chi t)^2}{1+M+B}\simeq\frac{F_Q/N}{1+M+B}.
\end{align}
This expression describes precisely how the sensitivity gain is limited by sub-optimal measurements ($M$) and noise ($B$). Depending on the chosen measurement scheme, we obtain $M_{\mathrm{L}}=N^4(\chi t)^6/6$, $M_{\mathrm{NL}}=N^6(\chi t)^{10}/270$, $M_{\mathrm{Q}}=N^8(\chi t)^{14}/875$ and $M_{\mathrm{MAI}}=0$. The effect of decoherence is accounted for by the terms $B_{\rm bal}=\epsilon N^{1+\gamma} (\chi t)^2$ and $B_{\rm dif}=\epsilon N\chi t$ for all measurements except MAI which gives rise to $B_{\rm bal,MAI}=4\epsilon N^{1+\gamma} (\chi t)^2$ and $B_{\rm dif}=2\epsilon N\chi t$ because of the two nonlinear evolutions of duration $\chi t$. The result~(\ref{eq:unifying}) predicts the scaling laws and optimal times in all cases discussed above. A detailed derivation of all results and an extended discussion will be exposed in a separate paper~\cite{LongPaper}.

\textit{Conclusions.---}We analytically identified the metrological potential associated with states that are generated by the one-axis-twisting evolution beyond the optimal squeezing time. We have found a unified scaling law of the metrological gain that continuously connects different optimal measurement strategies as a function of the preparation time. We further showed how the sensitivity scaling is modified by the presence of dominant dephasing processes in atomic experiments. For ballistic dephasing, we reveal in the thermodynamic limit a discontinuous behavior of the quantum gain as a function of the preparation time. The critical behavior presents characteristic traits of a phase transition. However, the discontinuity is observed here as a function of the evolution time in a dissipative system. Whether a connection with generalized concepts of quantum phase transitions~\cite{DiehlPRL2010,HeylREPPROG2018} can be established remains an interesting open question.

The critical preparation time, after which further quantum gains are suppressed by decoherence, predicts how much noise can be tolerated in order to sustain a quantum advantage with increasingly sensitive spin states. Our results identify optimal strategies for achieving significant quantum enhancements with non-Gaussian spin states in atomic experiments under realistic conditions.

\begin{acknowledgments}
\textit{Acknowledgments.---}A.S. and Y.B. acknowledge funding from the project macQsimal of the EU Quantum Flagship. M.G. acknowledges funding by the LabEx ENS-ICFP: ANR-10-LABX-0010 / ANR-10-IDEX-0001-02 PSL*.
\end{acknowledgments}


\begin{thebibliography}{}
\bibitem{footnoteQ} The maximal sensitivity for observables up to second order is obtained at an optimal time by the measurement of $\hat{X}_{\mathrm{Q}}=m_y \hat{S}_y+m_z \hat{S}_z +\frac{m_{xy}}{2} \{\hat{S}_x,\hat{S}_y\}+\frac{m_{xz}}{2} \{\hat{S}_x,\hat{S}_z\}$. At later times, this choice is no longer optimal among second-order observables but the sensitivity at the optimal time can not be surpassed. For this reason, we limit the second-order case to observables of this form.
\bibitem{footnoteSz}
For the states considered here, $(\langle \hat{S}_z^2\rangle)^{1/2}=\sqrt{N}/2$, and therefore the first-order approximation is valid for times $\chi t\ll 1/\sqrt{N}$. For the analysis of the MAI method below, however, we use the exact expression, Eq.~(\ref{eq:XMAI}).
\end{thebibliography}
\end{document}